# Neutron Scattering Study of the Spin Correlation in the Spin Ice System $Ho_2Ti_2O_7$

Masaki KANADA[1,2], Yukio YASUI[1,2], Yasuyuki KONDO[1], Satoshi IIKUBO[1], Masafumi ITO[1,2], Hiroshi HARASHINA[1,2], Masatoshi SATO[1,2], Hajime OKUMURA[3], Kazuhisa KAKURAI[2,3*] and Hiroaki KADOWAKI[4]

[1]*Department of Physics, Division of Material Science, Nagoya University, Furo-cho, Chikusa-ku, Nagoya 464-8602*

[2]*CREST, Japan Science and Technology Corporation (JST)*

[3]*Neutron Scattering Laboratory, ISSP, The University of Tokyo, Shirakata 106-1, Tokai, Ibaraki 319-1195*

[4]*Department of Physics, Tokyo Metropolitan University, Hachioji-shi, Tokyo 192-0397*

**Abstract**

Intensity distribution of the neutron magnetic scattering has been taken in the reciprocal space of a single crystal of the “spin ice” system $Ho_2Ti_2O_7$ at the elastic energy position in the temperature (*T*) region between 0.4 K and 50 K. The short range magnetic correlation becomes appreciable with decreasing *T* at around 30 K. The freezing of the moment system takes place at around 1.5 K with decreasing *T*. Based on the mean field treatment of the moment system, the distribution of the observed intensity has been reproduced, where the dipolar interaction is confirmed to be primarily important for the determination of the magnetic correlation. Simple calculations of the scattering intensity carried out by using a cluster of 25 moments of $Ho^{3+}$ ions indicate that there are many types of the moment arrangement within a cluster which have almost degenerated energies, indicating that the system cannot uniquely select the correlation pattern down to rather low temperature, and before going to the lowest energy state, the system undergoes the gradual slowing down to the glassy state.

corresponding author: M. Sato(e-mail address: e43247a@nucc.cc.nagoya-u.ac.jp)

*present address: Advance Science Research Center, JAERI, Tokai, Ibaraki 319-1195

## §1. Introduction

Pyrochlore compounds $R_2Ti_2O_7$ (R=rare earth elements, etc.) have a cubic structure[1] (space group $Fd\bar{3}m$) and consist of two kinds of 3-dimensional networks individually formed by the corner sharing $R_4$- and $Ti_4$-tetrahedra, respectively, as shown in Fig. 1. For R=Ho and Dy, it has been pointed out[2-4] that the magnetic moments of $R^{3+}$ ions have strong uniaxial anisotropy and the ferromagnetic nearest neighbor interaction. Because the magnetic easy axis at the corner site of a $R_4$ tetrahedron is parallel to the line which connects the site with the center of gravity of the tetrahedron, the ground state of four moments at the corners of a tetrahedron has the "two-in two-out" structure, where two moments direct inward and the other two direct outward. As can be easily known, this ground state has the six-fold degeneracy and therefore, the 3-dimensional network of these corner-sharing tetrahedra is expected, at least for systems with only nearest neighbor interaction, not to loose magnetic entropy even when the temperature approaches zero, that is, the moment system is highly frustrated. This situation is equivalent to that of solid water[5] in which the degeneracy associated with the arrangements of four hydrogen atoms around an oxygen atom site exists. Results of specific heat measurements have shown that the residual entropies of $Dy_2Ti_2O_7$[2] and $Ho_2Ti_2O_7$[6] are in good agreement with the predicted value for the solid water.[5] It has also been reported[7] that in $Ho_2Ti_2O_7$ no transition to a magnetically ordered state is found down to temperature $T \approx 46$ mK, even though the Weiss temperature $\Theta_{cw}$ estimated by fitting to the magnetic susceptibility-temperature curve is as large as ~+1 K. Hence, the moment systems of $R_2Ti_2O_7$ (R=Dy and Ho) are called "spin ice".

However, because the dominant interaction among the magnetic moments in $R_2Ti_2O_7$ (R=Dy and Ho) has been pointed out to be the dipole-dipole one[8-11] , which is rather long ranged and seems to reduce the degeneracy of the ground state, the problem may not be necessarily equivalent to that of solid water. In the present work neutron scattering study has been carried out on a single crystal of $Ho_2Ti_2O_7$ in the temperature region between 0.4 K and 50 K to clarify how the magnetic correlation grows as $T$ is lowered and the results are discussed mainly based on the mean field treatment and other simple calculations for a cluster of 25 moments.

## §2. Experiments

A single crystal of $Ho_2Ti_2O_7$ with a volume of ~ 0.8 $cm^3$ has been prepared by the floating zone (FZ) method. The magnetic susceptibility $\chi$ was measured by a SQUID magnetometer. From the results, the moment system has been found to have the ferromagnetic interaction and Weiss temperature $\Theta_{cw}$~1 K. Neutron measurements were carried out by using the triple-axis spectrometer at T1-1 of the thermal guide and C1-1 of the cold guide of JRR-3M of JAERI in Tokai. The 002 reflections of Pyrolytic graphite (PG) were used for both the monochromator and the analyzer. Horizontal collimations were 12'(effective)-open-60'-open at T1-1 and open-20'-20'-20' at C1-1, respectively. The condition of fixed initial neutron energy $E_i$ (=13.51 meV) was used at T1-1, where energy resolution is $\Delta E$ ~ 0.86 meV. At C1-1, the final neutron energy was fixed at 2.42 meV, where $\Delta E$ ~0.025 meV. A PG filter was introduced before the sample to suppress higher-order contamination. The crystal was oriented with its [110] axis vertical, where the points ($h$,$h$,$l$) in the reciprocal space can be reached.

## §3. Experimental Results and Discussion

Figure 2(a) shows the magnetic scattering intensity map taken at 0.4 K with the neutron transfer energy $E$= 0 meV by scanning the scattering vector $\boldsymbol{Q}$ in the ($h$,$h$,$l$) plane in the reciprocal space, with the T1-1 spectrometer, where the correction of the magnetic form factor $f(\boldsymbol{Q})$ has been made by using the numerical values of $Ho^{3+}$ ions calculated by Stassis *et al.*[12] The qualitative features of this map are consistent with those reported in ref. 13. In drawing the map, the nuclear Bragg peaks observed at $\boldsymbol{Q}$-points with even $h$ and $l$ or odd $h$ and $l$ and $\boldsymbol{Q}$-independent incoherent scattering are removed. Because the contribution of the nuclear incoherent scattering is found to be much smaller than that of the magnetic scattering, the ambiguity of the estimation of the nuclear incoherent scattering does not significantly affect the result. The magnetic scattering peaks were observed mainly around (0.8,0.8,0), (3.2,3.2,0), (0,0,1), (0,0,3) positions. The $\boldsymbol{Q}$-dependence of the scattering intensity at $E$= 0 meV becomes weak with increasing temperature and disappears completely at $T$ ~ 50 K. The $\boldsymbol{Q}$-dependence of the observed intensity along ($h$,$h$,0) and (0,0,$l$) at several temperatures are shown in Figs. 3(a) and 3(b), respectively.

We have also carried out the energy scan at $\boldsymbol{Q}$=(0,0,3) at 0.4 K up to the transfer energy of 7 meV and have not found inelastic scattering contribution, which is in clear contrast to the case of $Tb_2Ti_2O_7$.[14,15] Quasi elastic scattering contribution has not been observed, either, even under the condition of finer energy resolution of $\Delta E$ ~ 0.025 meV with the C1-1 spectrometer. These results are consistent with the results of Rosenkranz *et al.*,[16] who have determined the crystal-field parameters and the energy level scheme by measuring the crystal-field excitations in the wide energy region up to 90 meV and pointed out that ground state is a doublet with the strong uniaxial anisotropy.

In order to get rough information on the spatial correlation of the moment system, we have tried to calculate the magnetic scattering intensity $I(\boldsymbol{Q})$ by calculating the structure factor $\boldsymbol{S}(\boldsymbol{Q}) \propto \sum_j \boldsymbol{\mu}_j \times \exp(-i\boldsymbol{Q}\boldsymbol{r}_j)$ of clusters of 25 moments $\boldsymbol{\mu}_j$ ( j=1~25), which belong to eight tetrahedra shown in Fig. 4, and then by taking the average of $I(\boldsymbol{Q})$ over the equivalent correlation patterns which can be obtained by exchanging the $x$, $y$ and $z$ directions. In this expression of $\boldsymbol{S}(\boldsymbol{Q})$, $\boldsymbol{\mu}_{j\perp}$ is the perpendicular component of the $j$-th moment $\boldsymbol{\mu}_j$ to the scattering vector $\boldsymbol{Q}$ and $\boldsymbol{r}_j$ indicates the position of $\boldsymbol{\mu}_j$ . The condition of "two-in two-out" structure is considered in the calculation( Because the form factor correction is made for the observed data, we have not included the form factor in the expression.). Then, we have found that several patterns can roughly reproduce the observed $\boldsymbol{Q}$-dependence of the scattering intensity. The thin arrows in Fig. 4 show one of these correlation patterns, for example.

Because the dipole interaction among the magnetic moments is dominant in this system, we have also calculated the dipole energy of the 25 moments for all possible correlation patterns within the restriction of "two-in two-out" structure. From the results, the dipole energies have been found to be roughly degenerate. If one of 25 moments turns its direction and the "two-in two-out" structure is broken, the dipole energy of the 25 moments increases by ~5 K. We can therefore expect at $T$= 0.4 K, that each tetrahedron has the "two-in two-out" structure. Even with this restriction, there are many choices of the moment arrangements, depending on what direction the net moment of each tetrahedron directs. The dipole energy of these correlation patterns are degenerate within the energy of ~0.5 K. As an example of the calculated intensity maps, the result for the correlation pattern shown in Fig. 4 is displayed in Fig. 2(b). There is, however, no reason that it is uniquely realized at low temperature. Instead, we have to consider the state which is described by the coexistence of other local correlation patterns with almost degenerate energies and possibly with similar $\boldsymbol{S}(\boldsymbol{Q})$.

In order to discuss the $T$-dependence of the scattering intensity, we introduce the decay factor α of the spatial correlation of the moments as shown in the inset of Fig. 3(a), where the decay is phenomenologically taken into account as the contraction of the effective absolute value of the moments $\mu$ by the factor α (<1). In the figure, the "nearest" sites indicate the ones which can be reached from the relevant site through an edge of the tetrahedra, and the "next nearest" sites indicate the ones which can be reached through two edges of the tetrahedra. If α=1, the system is fully correlated. As $T$ increases, α is expected to decrease. At each temperature studied here, we have chosen the α value and the background count to reasonably explain the observed $\boldsymbol{Q}$-dependence of $\boldsymbol{S}(\boldsymbol{Q})$ along ($h$,$h$,0) and (0,0,$l$), simultaneously. The results are shown in Figs. 3(a) and 3(b) by the solid lines for the correlation pattern shown in Fig. 4. The values of α are 0.9 at 0.4 K, 0.7 at 1.7 K, 0.6 at 6 K and 0.0 at 50 K, respectively, Because even at 6 K, $\alpha^2$ is as large as ~ 0.36, we cannot, strictly speaking, ignore the effect of the magnetic moments at the sites further than the "next nearest" neighbor ones. However, we find that the fitted results can qualitatively reproduce experimental results.

We have also adopted the mean field analysis of the experimental results.

Because the details of the calculations can be found elsewhere,[17)] here we describe them very briefly. First, the spin Hamiltonian is written in the quadratic form as

$$H = -\sum_{n,\upsilon,\alpha,n',\upsilon',\beta} J_{n,\upsilon,\alpha;n',\upsilon',\beta} S_{n,\upsilon,\alpha} S_{n',\upsilon',\beta} \quad , \tag{1}$$

where $\boldsymbol{S}_{n,\nu,\alpha}$ represents α-component (α=$x$,$y$,$z$) of a classical vector spin $\boldsymbol{S}_{n,\nu}$=$\boldsymbol{S}_{\boldsymbol{t}n+\boldsymbol{d}\nu}$($|\boldsymbol{S}_{n\nu}|$=1 and the Ho magnetic moment is described as $\mu\boldsymbol{S}_{n,\nu}$ with $\mu\cong 10\ \mu_B$) on the ν-th site in the $n$-th unit cell located at $\boldsymbol{t}_n$(Here, we use the primitive unit cell with four Ho sites at $\boldsymbol{d}_\nu$(ν=1,2,3,4), instead of the face-centered cubic cell). The detailed Hamiltonian is composed of three kinds of energies, the single ion anisotropy energy and the exchange and dipole-dipole interaction terms as shown in this order in the right hand side of the following equation.

$$\begin{aligned} H = & -D_a \sum_{n,\upsilon} [(n_\upsilon \cdot S_{n,\upsilon})^2 - |S_{n,\upsilon}|^2] - J_1 \sum_{\langle n,\upsilon;n'\upsilon'\rangle} S_{n,\upsilon} \cdot S_{n',\upsilon'} \\ & + D_{dp} r_{nn}^3 \sum_{\langle n,\upsilon;n'\upsilon'\rangle} \left[\frac{S_{n,\upsilon} \cdot S_{n',\upsilon'}}{|r_{n,\upsilon;n',\upsilon'}|^3} - \frac{3(S_{n,\upsilon} \cdot r_{n,\upsilon;n',\upsilon'})(S_{n',\upsilon'} \cdot r_{n,\upsilon;n',\upsilon'})}{|r_{n,\upsilon;n',\upsilon'}|^5}\right], \end{aligned} \tag{2}$$

where, $\boldsymbol{r}_{n,\nu;n',\nu'}$=$\boldsymbol{t}_n$+$\boldsymbol{d}_\nu$-$\mathbf{t}_{n'}$-$\boldsymbol{d}_{\nu'}$ and other notations are as follows. $D_a$ is the anisotropy parameter. The vector $\boldsymbol{n}_\nu$($|\boldsymbol{n}_\nu|$=1) indicates the local easy-axis direction. Only the nearest-neighbor exchange interaction with the coupling constant $J_1$ is considered in the second term(It can be justified, because in the present system the dipole interaction plays, as shown later, a dominant role.). The third term is the dipolar interaction among the Ho magnetic moments $\mu\boldsymbol{S}_{n,\nu}$. The dipole interaction constant is described as $D_{dp}$=$\mu^2/r_{nn}^3$ ,where $r_{nn}$ is the distance between nearest neighbor moments. Then, the Fourier transform of the interaction constant $J_{n,\upsilon,\alpha;n',\upsilon',\beta}$ is calculated as

$$J_{q;\upsilon,\alpha;\upsilon',\beta} = \sum_n J_{n,\upsilon,\alpha;n',\upsilon',\beta} \exp[-iq \cdot (t_n + d_\upsilon - t_{n'} - d_{\upsilon'})] \quad , \tag{3}$$

where $\boldsymbol{q}$ is the wave vector in the first Brillouin zone. These components form the 12×12 matrix with suffices ν (=1-4) and α or β(=$x$, $y$ and $z$), for which eigenvalue equations are expressed[18)] as

$$\sum_{\upsilon',\beta} J_{q;\upsilon,\alpha;\upsilon',\beta} u_{q;\upsilon',\beta}^{(\rho)} = \lambda_q^{(\rho)} u_{q;\upsilon,\alpha}^{(\rho)} \quad . \tag{4}$$

The eigenvalues $\lambda_{\boldsymbol{q}}^{(\rho)}$(ρ=1,2...12) and the eigenvectors $u_{\boldsymbol{q};\nu,\alpha}^{(\rho)}$ were calculated at fixed $\boldsymbol{q}$ by numerical diagonalization of the 12-dimensional real symmetric-matrix $J_{\boldsymbol{q};\nu,\alpha;\nu',\beta}$ with the normalization condition

The eigenvalues $\lambda_{\boldsymbol{q}}^{(\rho)}(\rho=1,2...12)$ and the eigenvectors $u_{\boldsymbol{q};\,\nu,\,\alpha}^{(\rho)}$ were calculated at fixed $\boldsymbol{q}$ by numerical diagonalization of the 12-dimensional real symmetric-matrix $J_{\boldsymbol{q};\,\nu,\,\alpha;\nu',\beta}$ with the normalization condition

$$\sum_{\upsilon,\alpha} u_{q;\upsilon,\alpha}^{(\rho)} u_{q;\upsilon,\alpha}^{(\sigma)*} = \delta_{\rho,\sigma} \quad . \tag{5}$$

by this eigenvector $u_{\boldsymbol{q};\nu,\alpha}^{(\rho)}$, the moment distribution $S_{n;\nu,\alpha}$ of the eigenmode, which responds to the magnetic field with the wave vector $\boldsymbol{q}$, is given by

$$S_{n;\nu,\alpha} = S_{\boldsymbol{q}0}^{(\rho)} u_{\boldsymbol{q};\nu,\alpha}^{(\rho)} \exp[i\boldsymbol{q}(\boldsymbol{t}_n+\boldsymbol{d}_\nu)] + \text{c.c.}, \tag{6}$$

where $S_{\boldsymbol{q}0}^{(\rho)}$ is the amplitude of the modulation. Because each mode is decoupled from the other ones, it is easy to deduce the wave vector dependent susceptibility $\chi_{\boldsymbol{q}}$ and neutron scattering intensity $I(\boldsymbol{Q}=\boldsymbol{G}+\boldsymbol{q})$ ($\boldsymbol{G}$ : reciprocal lattice vector) as shown by the following equations[17,19,20] for the present non-Bravais lattice by the mean field treatment.

$$\chi_{\boldsymbol{q};\nu,\alpha;\nu',\beta} \propto \sum_{\sigma} \frac{u_{q;\upsilon,\alpha}^{(\sigma)} u_{q;\upsilon',\beta}^{(\sigma)*}}{3k_B T - 2\lambda_q^{(\sigma)}} \tag{7}$$

and

$$I(Q = G + q) \propto f(Q)^2 k_B T \sum_{\alpha,\beta,\upsilon,\upsilon'} (\delta_{\alpha\beta} - \hat{Q}_\alpha \hat{Q}_\beta) \sum_{\sigma} \frac{u_{q;\upsilon,\alpha}^{(\sigma)} u_{q;\upsilon',\beta}^{(\sigma)*}}{3k_B T - 2\lambda_q^{(\sigma)}} \cos[G \cdot (d_\upsilon - d_{\upsilon'})], \tag{8}$$

where $f(\boldsymbol{Q})$ is the magnetic form factor of the moments. From eq. (7), it is easily found that the Curie temperature $T_C^{MF}$ predicted by this theory can be expressed by using the maximum value of the eigenvalues, $\lambda_{\boldsymbol{q}m}$ as $T_C^{MF}=2\lambda_{\boldsymbol{q}m}/3k_B$. (In the actual system, the ferromagnetic ordering does not take place at least down to ~46 mK[7], possibly because of the existence of the strong frustration.)

Here, to reproduce the strong local Ising anisotropy, $D_a$~ 810 K is adopted, and the dipole interaction $D_{dp}$ is estimated to be 1.4 K for the $r_{nn}$ value of the present system. The sum of the dipole interaction is taken for the moments within a sphere with the radius of ~100 A. We use $I(\boldsymbol{Q})/f(\boldsymbol{Q})^2$ as a trial function, and perform the least square fitting to the form-factor-corrected intensity observed in the present experiment with the temperature, the exchange constant $J_1$ and the scale factor being the fitting parameters. (Here, $T$ is treated as a parameter, by assuming that we can find "effective" value of $T$ to describe the $\boldsymbol{Q}$-dependence of the magnetic correlation reasonably well even in the $T$ region much lower than $T_C^{MF}$. We argue this point once more later.)

observed results can be reproduced almost equally well for the "effective" temperature of $T_{=}1.9$ K, (as shown in Fig. 2(c) and Figs. 5(a) and 5(b)), and because by the fitting with $D_{\mathrm{dp}}$ being fixed at 0, the observed map can hardly be reproduced, it can be known that under the existence of the large single ion anisotropy $D_{\mathrm{a}}$, the role of the dipole interaction is much more important than that of the exchange interaction. The similar results have been obtained for a powder sample of $Ho_2Sn_2O_7$.[17)]

Now, we have shown that the present mean field treatment can reproduce the characteristics of the intensity distribution of the magnetic scattering rather well in the whole area of the reciprocal space experimentally studied here. This seems to empirically support the use of the "effective" temperature to study the $\boldsymbol{Q}$-dependence of the intensity or the correlation pattern of the highly frustrated moment system, and the introduction of this kind of temperature may be considered to have a role to renormalize the value of $\lambda_q^{(\sigma)}$ (see eqs. (7) and (8)). However, its theoretical justification has to be made. At this moment, results of the present analyses indicate, at least, that the dominant interaction among the moments is the dipolar one

Figure 6 shows the magnetic structure expected by the mean field theory to be realized at $T_{\mathrm{C}}^{\mathrm{MF}}$, which corresponds to the eigenvalue $\lambda_{\boldsymbol{q}\mathrm{m}}$. In the actual system, the ordering does not take place and the freezing of the moments appears, when $T$ is lowered, at the temperature $T_{\mathrm{f}}$ ~1.5 K as is mentioned below.

Figure 7 shows the $T$-dependence of the intensity at the $\boldsymbol{Q}$ points, (0,0,3) and (0.9,0.9,3) where the peak and the dip of the scattering intensity are observed, respectively. The $\boldsymbol{Q}$-dependence becomes appreciable with decreasing $T$ below ~ 30 K. It suggests that the short range correlation begins to grow at around this temperature with decreasing $T$. Inset shows the hysteretic behavior observed below 1.5 K with the C1-1 spectrometer by changing the temperature up and down. The time dependence of the scattering intensity at $\boldsymbol{Q}$=(0,0,1) has also been observed at 0.4 K after cooling to the temperature, as shown in Fig. 8, where the typical time scale of the intensity change is of the order of one hundred minutes. These results indicate that the moment system begins to freeze at around ~ 1.5 K with decreasing $T$, before exhibiting the long range order even though it is expected theoretically.[9)]

Based on the results of the present studies, we think that the magnetic correlation in the present system can be understood as follows. The short range correlation begins to grow at around 30 K with decreasing temperature, and the observed intensity distribution shown in Fig. 2(a) is realized at low temperatures. However, because there are many correlation patterns whose energies are almost degenerate within the energy difference of ~0.5 K (The differences of the dipole energies among the correlation patterns of 25 moments (see Fig. 4) which roughly reproduce the observed $\boldsymbol{Q}$-dependence of $I(\boldsymbol{Q})$ are of the order of 0.5 K), all these patterns may exist down to $T$~0.5

K. Then, before going to the critical temperature, at which the long range magnetic ordering expected by the Monte-Carlo calculation[9)] takes place, the moment clusters undergo the freezing to the glassy state as observed at 1.5 K.

Figure captions

Fig. 1. Left figure shows the three-dimensional networks of $Ho_4$- and $Ti_4$-tetrahedra in the structure of $Ho_2Ti_2O_7$. The right figure shows the Ho moments at the corner of a tetrahedron which have the “two-spin-in two-spin-out” structure.

Fig. 2. Maps of the scattered neutron intensity taken with the spectrometer setting at $E$=0 meV are shown in ($h$,$h$,$l$) plane: (a) observed data(form factor corrected) and (b) calculated map corresponding to the moment cluster shown in Fig. 4. The map in (c) shows the result of the mean field analysis. (Note that none of quasi elastic and inelastic scattering has been observed up to 7 meV for the present system.)

Fig. 3. ***Q***-dependence of the form factor corrected magnetic scattering intensity is shown along (a) ($h$,$h$,0) and (b) (0,0,$l$). Solid lines show results of the intensity calculations for clusters of 25 moments at the corners of 8 tetrahedra shown in Fig. 4. In the calculation, the spatial correlation of the moments is considered by expressing their magnitudes as shown in the inset of 3(a) for the center sites and its nearest and the next nearest neighbor sites(Here, the “next nearest” indicates the sites which can be reached from the center site through two edges of the tetrahedra). See text for details.

Fig. 4. Example of the magnetic correlation patterns which can roughly reproduce the observed intensity map shown in Fig. 2(a). The thick arrows show the net moment directions of the corresponding $Ho^{3+}$ tetrahedra. The thin arrows show the directions of the magnetic moments of $Ho^{3+}$ ions along the lines which connect their sites with the center of gravity of the tetrahedra. Only the tetrahedra which contain a $Ho^{3+}$ site nearest from the one at the center position of the cube shown in the figure are drawn. The 25 moments at the corners of these tetrahedra are used in the model calculations.

Fig. 5. Results of the fitting by the mean field analysis to the data in Fig. 2(a) are shown along (a) ($h$,$h$,0) and (b) (0,0,l) at several temperatures.

Fig. 6. Magnetic structure of $Ho_2Ti_2O_7$ expected by the mean field analysis at its critical temperature is drawn in the similar way to the case of Fig. 4.

Fig. 7. $T$-dependence of the scattered neutron intensities observed with the T1-1 spectrometer at the ***Q*** points (0,0,3) and (0.9,0.9,3), where the peak and the bottom, respectively, of the intensity are observed. Inset shows the hysteretic behavior observed with C1-1 spectrometer at (0,0,1) by scanning $T$ up and down. They have been taken at $E$=0 meV.

Fig. 8. Scattering intensity observed with the transfer energy $E$=0 meV at ***Q***=(0,0,1) after cooling down to 0.4 K is shown as a function of the elapsed time.

$Ho_2Ti_2O_7$

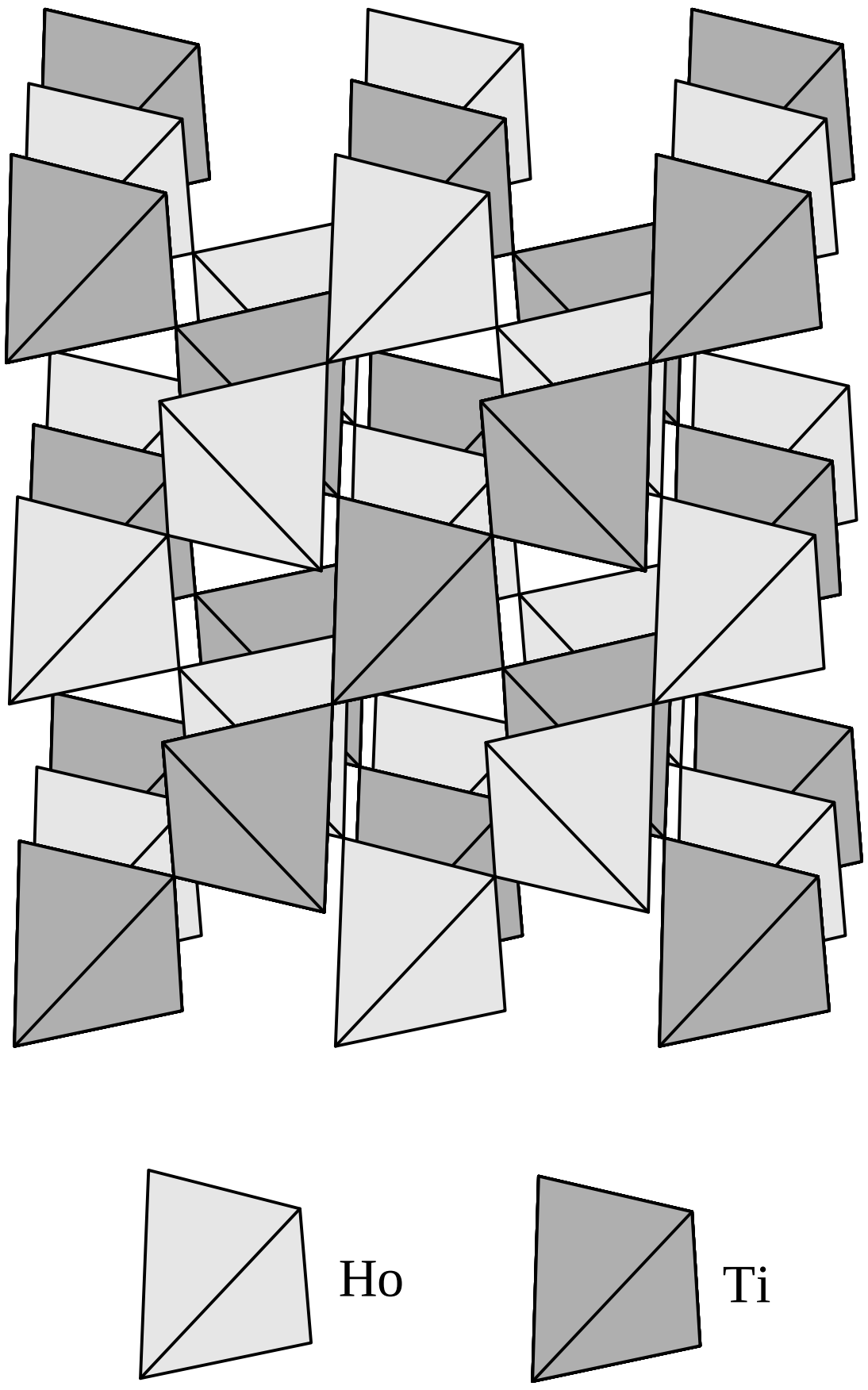

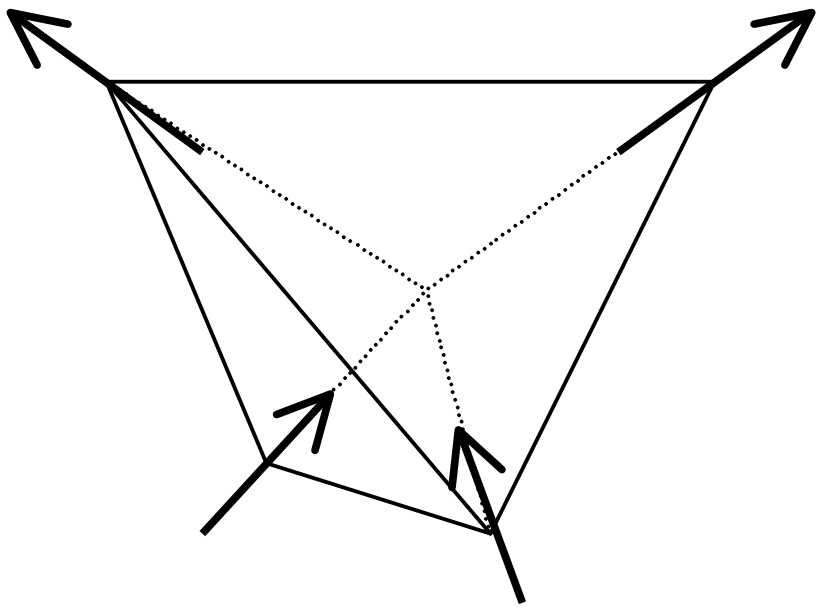

**Fig.1**

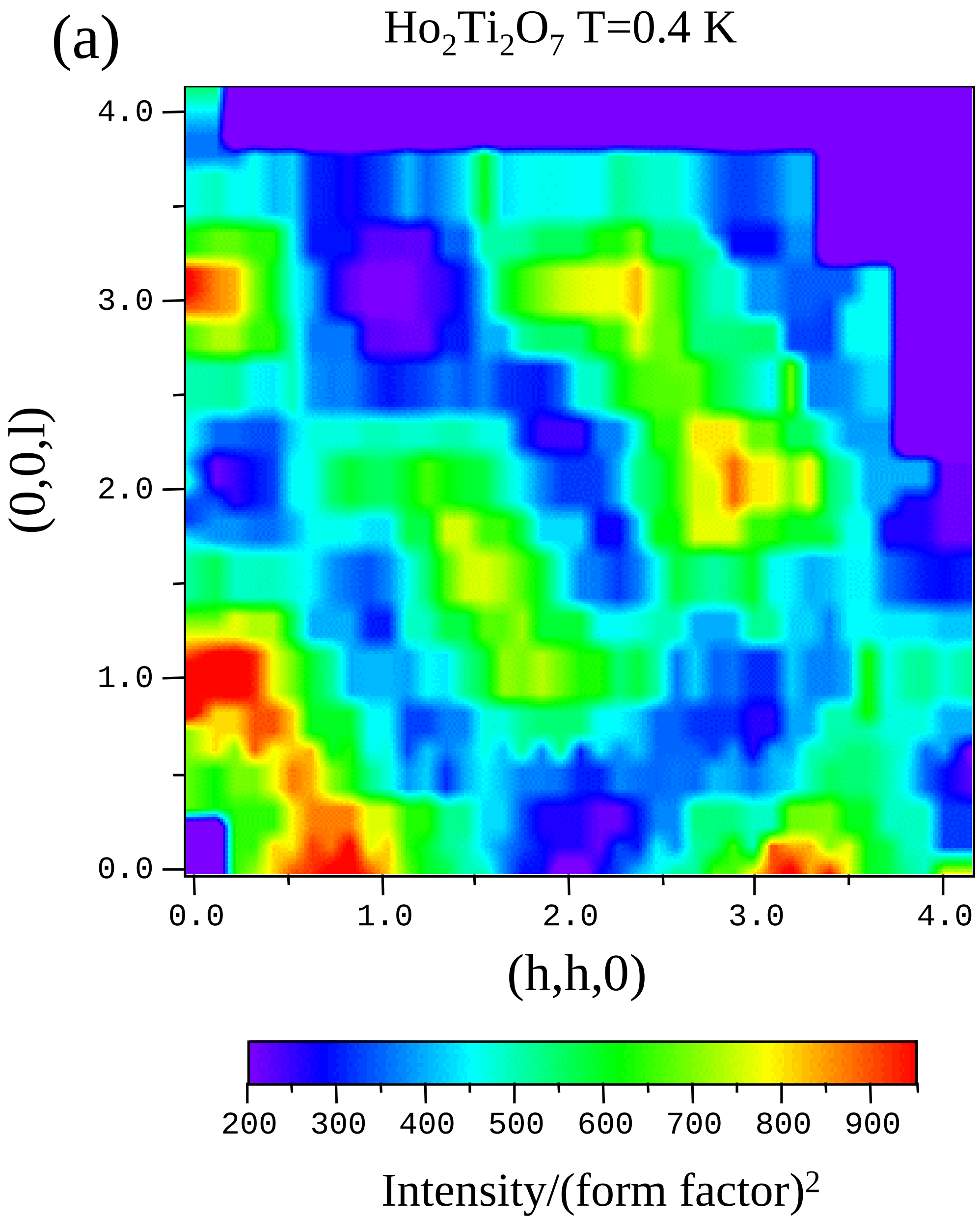

**Fig.2(a)**

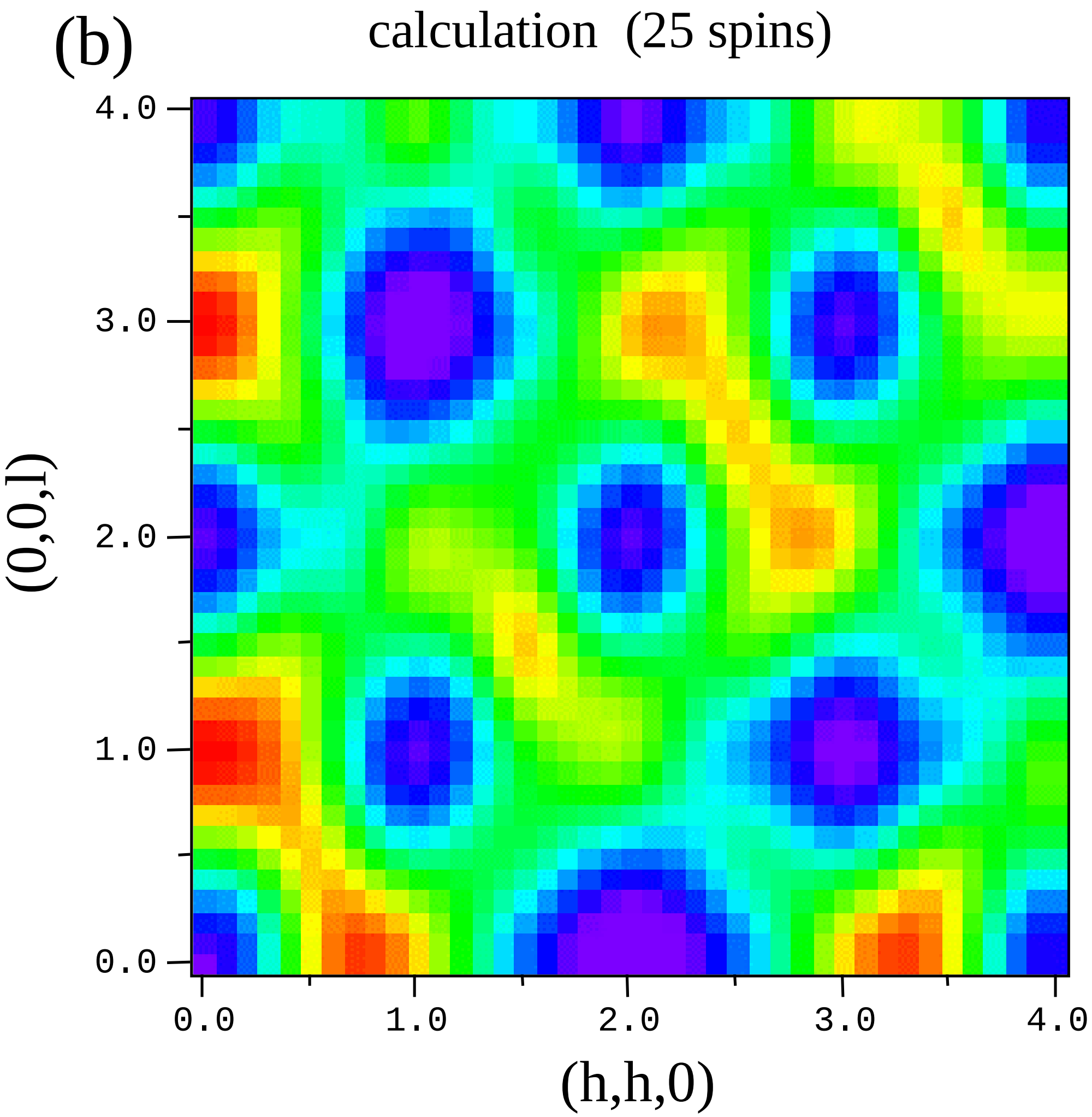

**Fig.2(b)**

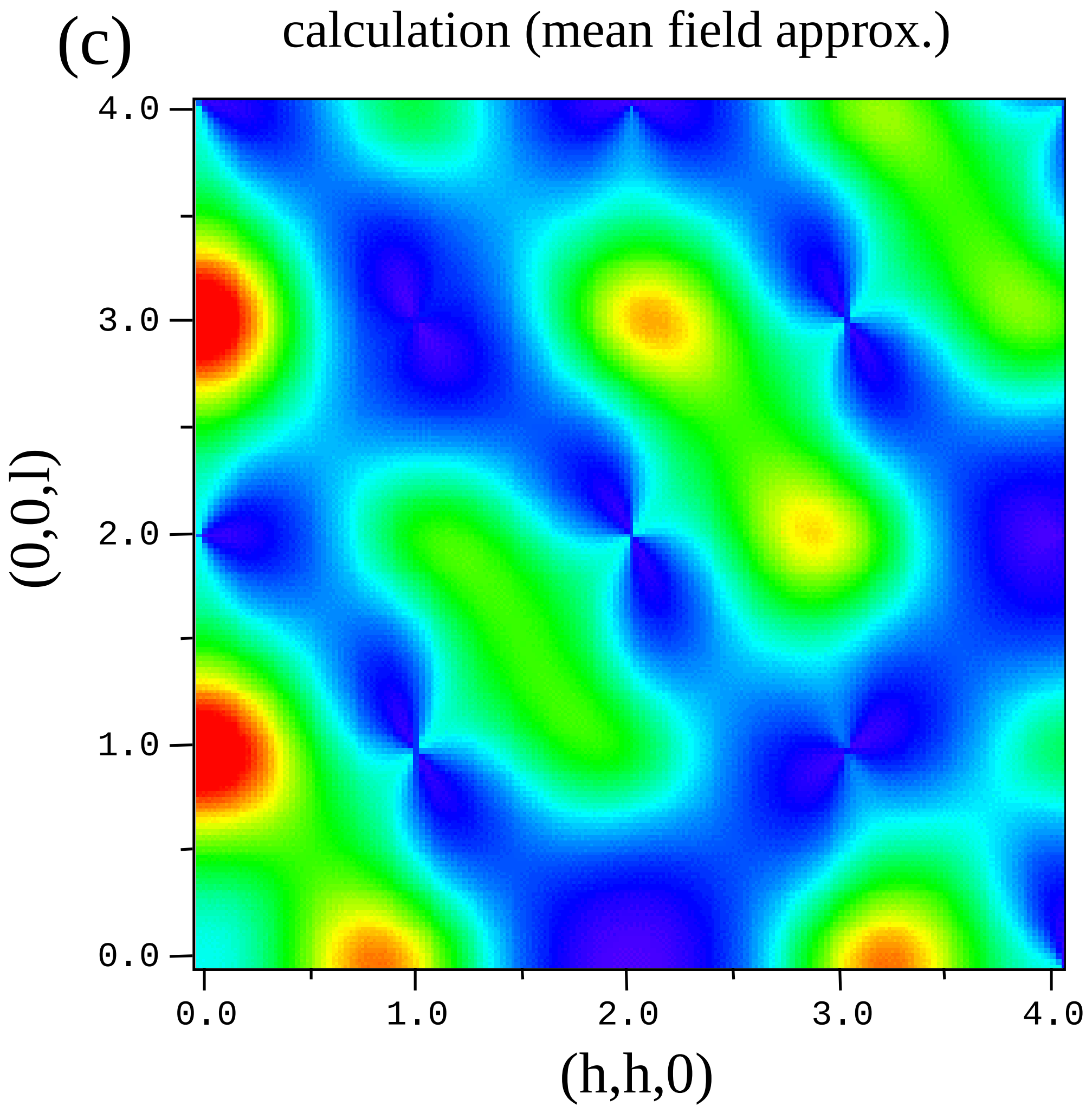

**Fig.2(c)**

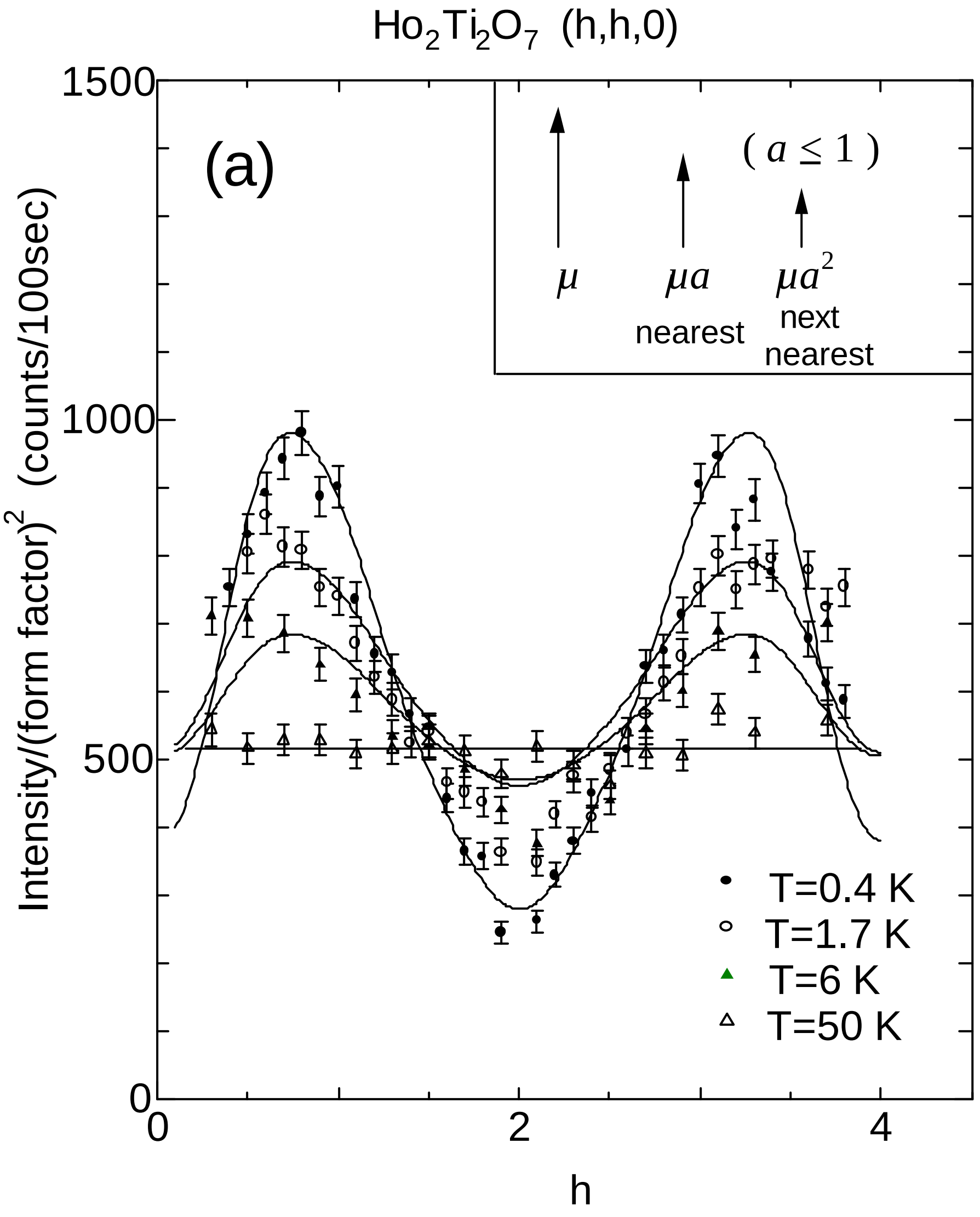

**Fig. 3(a)**

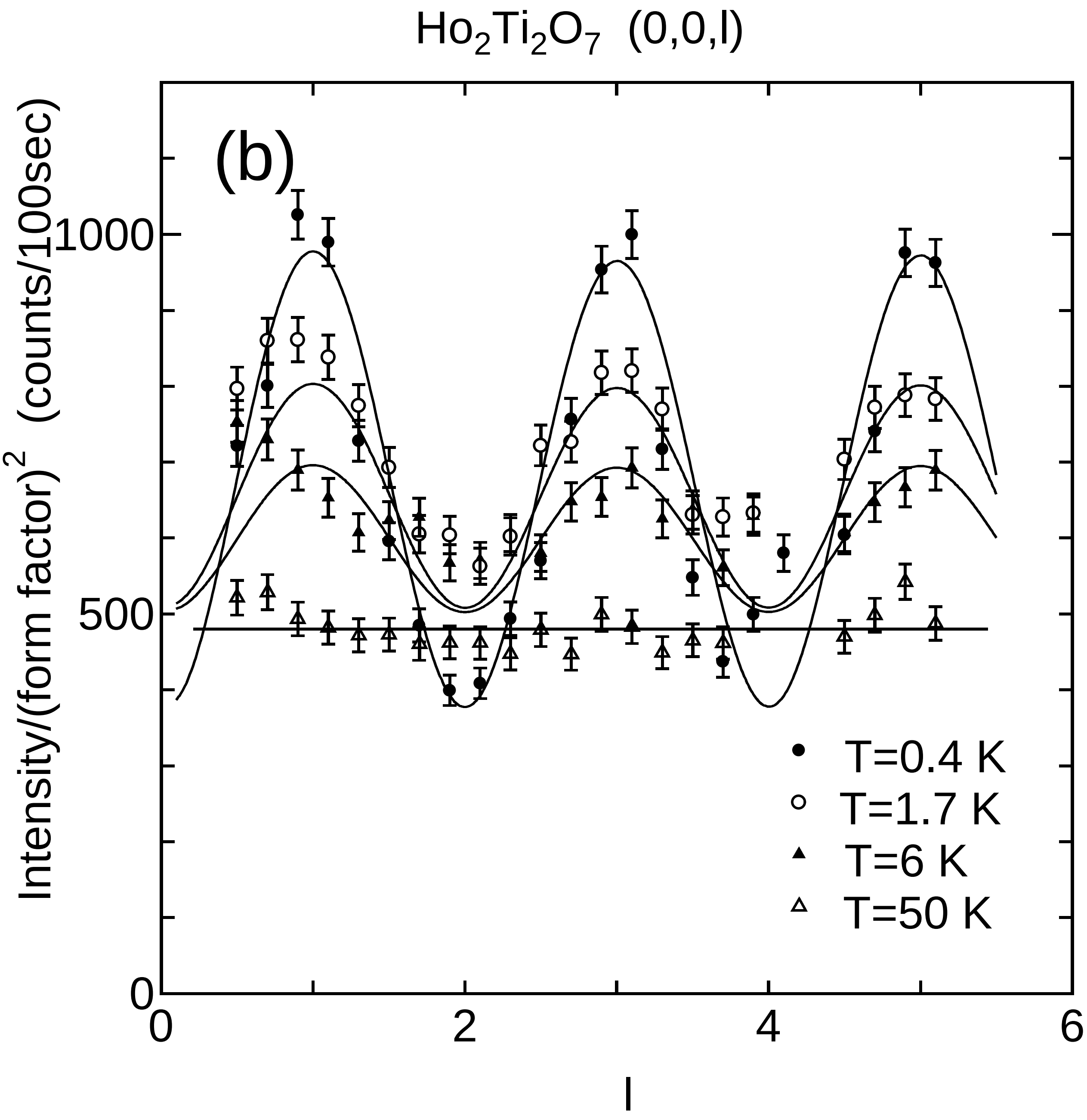

**Fig.3(b)**

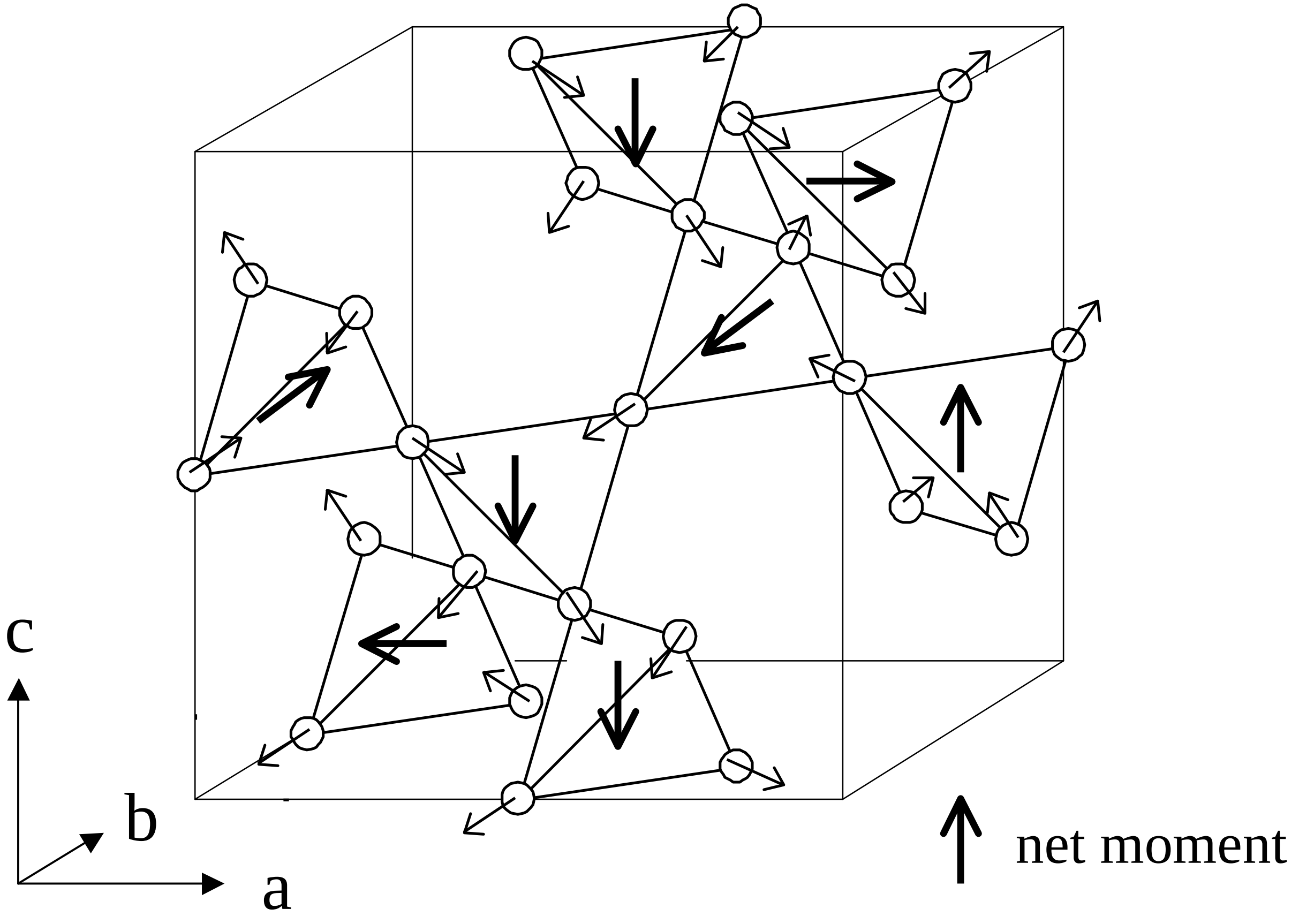

**Fig.4**

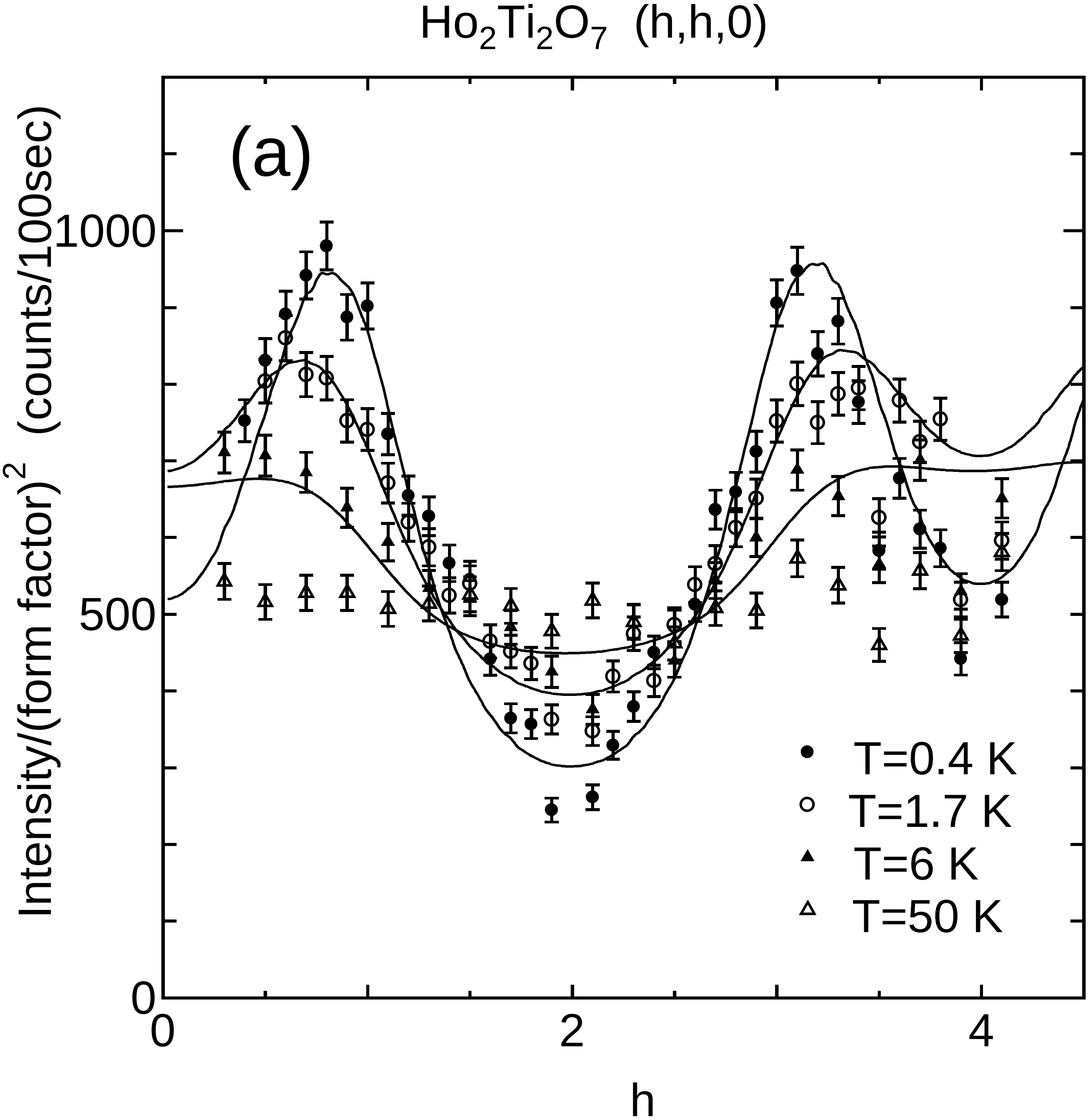

**Fig.5(a)**

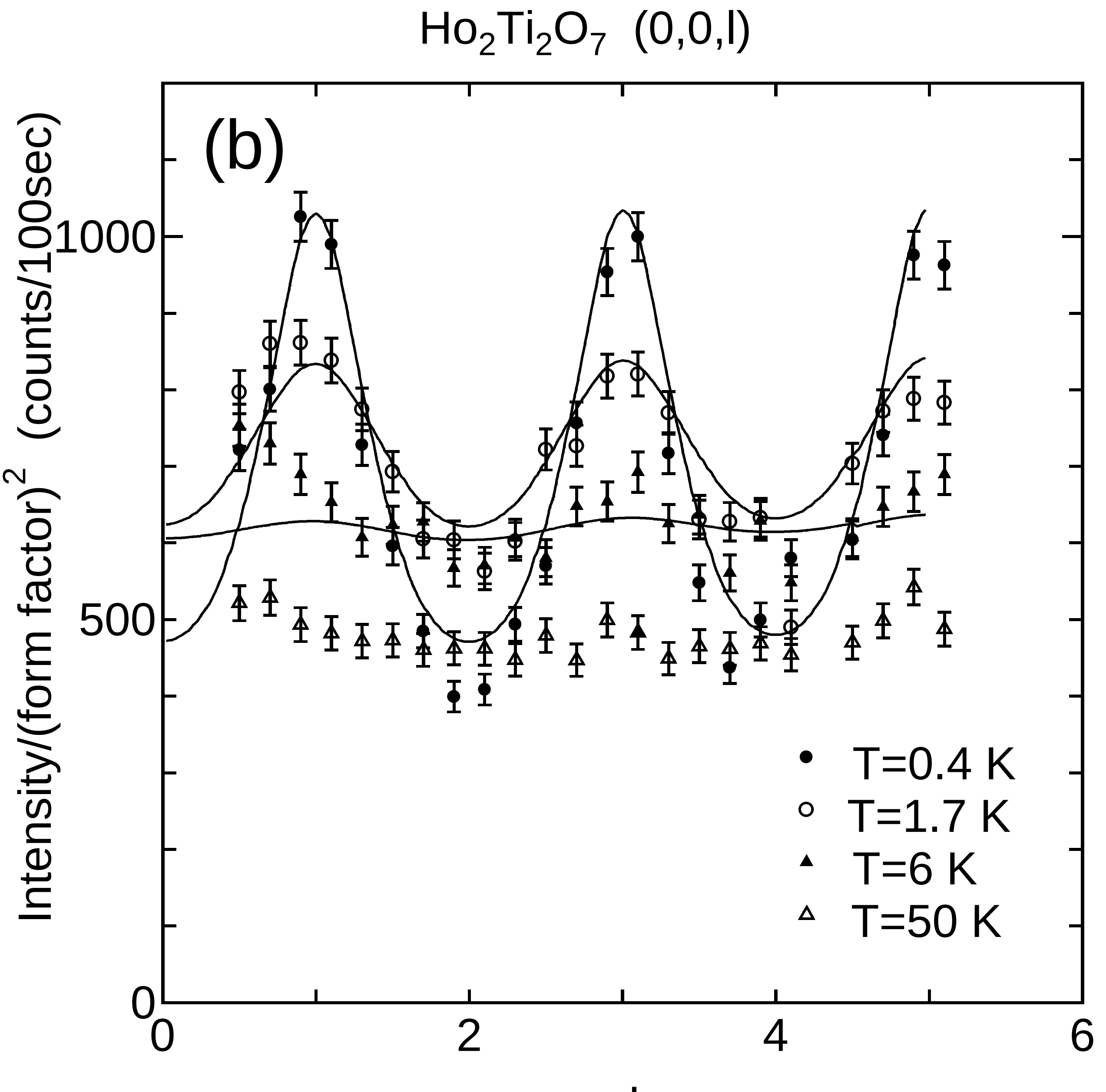

**Fig.5(b)**

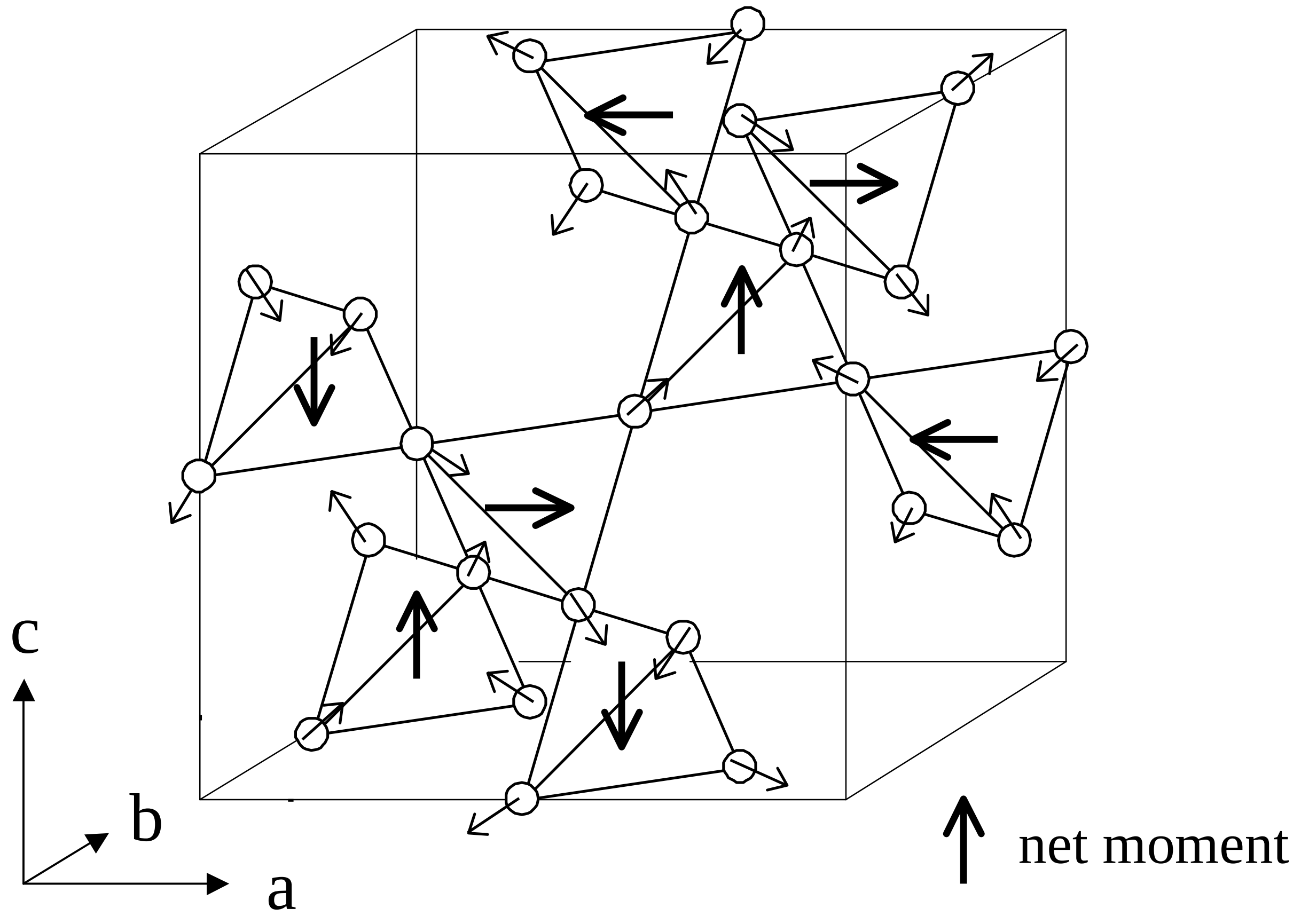

**Fig.6**

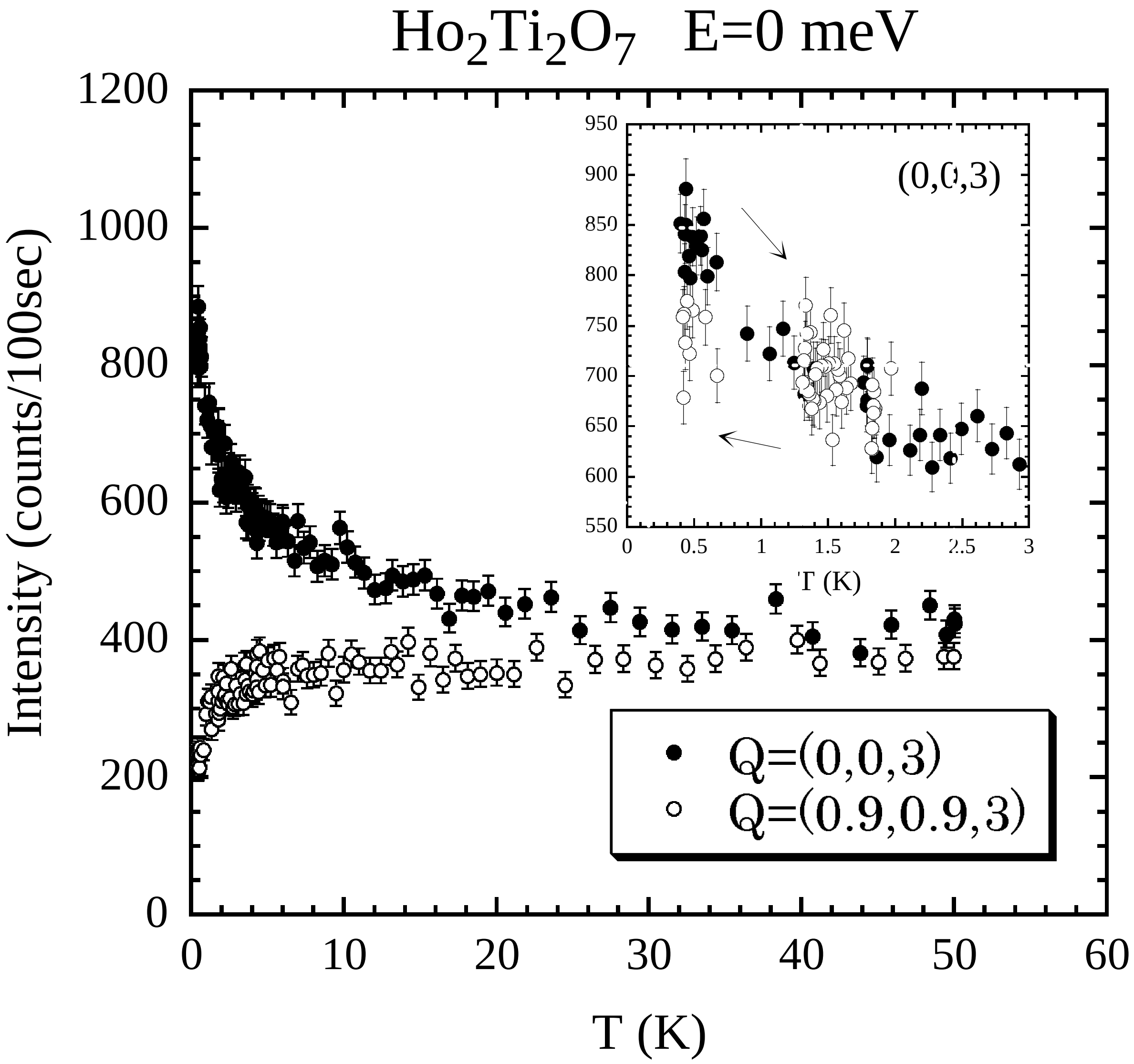

**Fig.7**

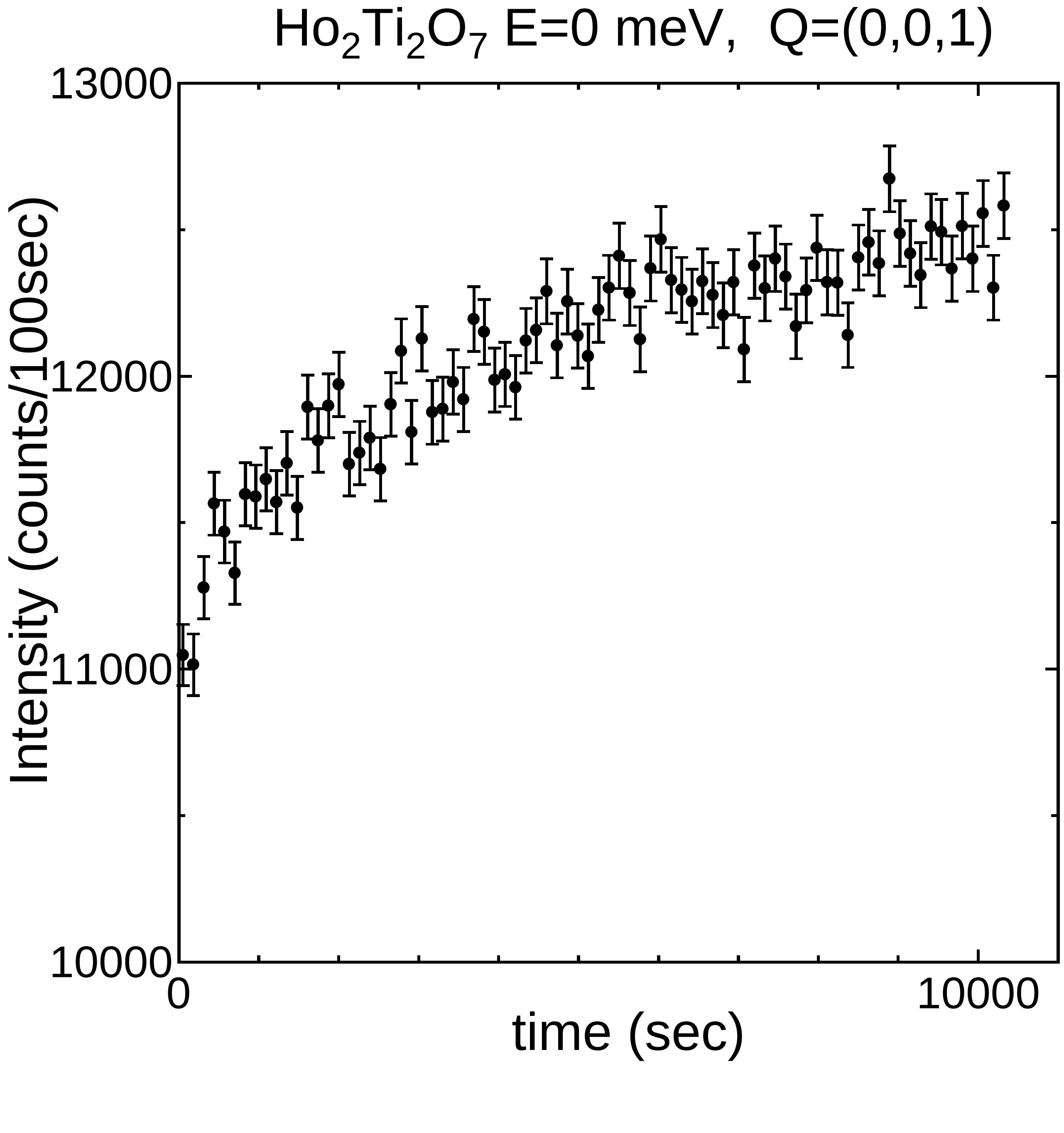

**Fig.8**